\documentclass[preprint, aps, draft]{revtex4}

\begin{document}
\title{From information to quanta: A derivation of the geometric formulation of quantum theory from information geometry}

\author{Marcel Reginatto}
\affiliation{Physikalisch-Technische Bundesanstalt, Bundesallee 100,
38116 Braunschweig, Germany}

\begin{abstract}

It is shown that the geometry of quantum theory can be derived from geometrical structure that may be considered more fundamental. The basic elements of this reconstruction of quantum theory are the natural metric on the space of probabilities (information geometry), the description of dynamics using a Hamiltonian formalism (symplectic geometry), and requirements of consistency (K\"{a}hler geometry). The theory that results is standard quantum mechanics, but in a geometrical formulation that includes also a particular case of a family of nonlinear gauge transformations introduced by Doebner and Goldin. The analysis is carried out for the case of discrete quantum mechanics. The work presented here relies heavily on, and extends, previous work done in collaboration with M. J. W. Hall.

\end{abstract}

\maketitle

\section{Introduction}

It is now known that quantum mechanics has a rich geometrical structure which allows for a geometric formulation of the theory. The geometric approach was introduced by Kibble \cite{K79} and later further developed by a number of other authors. One successful strategy has been to start from a known formulation of quantum mechanics and to identify geometrical features that can be used for the reformulation of the theory. This paper inverts this procedure: the aim is to {\it derive} the geometry of quantum theory from geometrical structure that may be considered more fundamental, and to examine the assumptions that are needed to do this. The work presented here relies heavily on, and extends, previous work of Reginatto and Hall \cite{RH12,RH13}. The basic elements of this geometrical reconstruction of quantum theory are the natural metric on the space of probabilities (information geometry), the description of dynamics using a Hamiltonian formalism (symplectic geometry), and requirements of consistency (K\"{a}hler geometry). One may associate a Hilbert space with the K\"{a}hler space, which leads to the standard version of quantum theory. The analysis is carried out for the case of discrete quantum mechanics; a similar approach has been carried out previously for continuous systems \cite{RH12}.

\section{Information geometry}

Consider a system with a discrete configuration space. If the configuration of the system is subject to uncertainty, the state of the system will be described by a probability $P=(P^1,...,P^n)$ where $n$ is the number of states. The probability that the system is in state $i$ is $P^i$, where $P^i$ satisfies $P^i \geq 0$ and $\sum_i P^i=1$. The space of probabilities can be visualized as points in the simplex $S_{n-1} = \{P \in \textbf{R}_n^+:\sum P^i=1\}$.

There is a natural line element in this space, given by
\begin{equation}\label{naturalLineElement}
ds^2 = G_{ij} \; dP^i \; dP^j  = \frac{\alpha}{2P^i} \; \delta_{ij} \; dP^i \; dP^j,
\end{equation}
where $\alpha$ is a constant. The value of $\alpha$ can not be determined {\it a priori}; it is usually set to $\frac{1}{2}$, but for the purposes of this paper it will be convenient not to follow this convention. Instead, $\alpha$ will be treated as a parameter. The line element of Eq. (\ref{naturalLineElement}) defines a distance on the space of probabilities. This distance seems to have been introduced into statistics by Bhattacharyya \cite{B43, B46} as a way of providing a measure of divergence for multinomial probabilities \cite{G90}. Wootters has shown that this distance can be motivated using arguments based on the concept of the distinguishability of states and refers to this distance as the ``statistical distance'' \cite{W81}.

The metric $G_{ij}$ that appears in Eq. (\ref{naturalLineElement}) is known as the \emph{information metric},
\begin{equation}\label{metricP}
G_{ij} = \frac{\alpha}{2P^i} \; \delta_{ij}.
\end{equation}
It has been shown by \v{C}encov that the information metric is the only metric that is invariant under a family of probabilistically natural mappings known as {\it congruent embeddings by a Markov mapping} \cite{C81}. A simpler proof, which also makes use of these mappings, was given later by Campbell \cite{C86}. It will be useful to give a brief description of these mappings because their generalization (see below) play an important role in the derivation of the main result of this paper.

I follow the presentation of Ref. \cite{C86}. A Markov mapping is a particular type of linear transformation between a simplex $S_{m-1}$ and a simplex $S_{n-1}$ (with $m \leq n$) which preserves the probability; i.e., $\sum_{a=1}^m P^a = \sum_{b=1}^n \tilde{P}^b = 1$. For $m=n$, the mapping is just a permutation of the components $P^i$, but for $m<n$, the mapping relates spaces of different dimensions. A Markov mapping may be constructed in the following way. Let $A=\{A_1,...,A_m\}$ be a partition of the set $\{1,2,...,n\}$ into disjoint sets. Associate a probability vector $Q_{(a)}=(q_{a1},...,q_{an})$ to each of the $A_a$, where the $q_{ab}$ satisfy
\begin{equation}
q_{ab} = 0 \texttt{ if } b \notin A_a, ~~~~~~~~~~q_{ab} > 0 \texttt{ if } b \in A_a, ~~~~~~~~~~\sum_{b=1}^n q_{ab} = 1.
\end{equation}
The probability vector $Q_{(a)}$ is therefore concentrated on $A_a$. Note that the $m \times n$ matrix $\textbf{Q}$ with elements $q_{ab}$ has the following properties: Each column has precisely one non-zero element and each row sums to one.

Define mappings between $f:S_{m-1} \rightarrow S_{n-1}$ and $g: S_{n-1} \rightarrow S_{m-1}$ by
\begin{eqnarray}
\tilde{P}^b &=& \sum_{a=1}^m P^a q_{ab},\nonumber\\
P^a &=& \sum_{b \in A_a} \tilde{P}^b,~~~~~~~~a \in \{1,2,...,m\}.
\end{eqnarray}
Following \v{C}encov, the mapping $f$ is known as a congruent embedding of $S_{m-1}$ in $S_{n-1}$ by a Markov mapping. The mapping $g$, which is also defined in terms of the partition $A$, has the property that the composition $g \circ f$ is the identity map on $S_{m-1}$.

I now consider a simple example of a Markov mapping for the case where $n=m+1$. Set
\begin{equation}
\textbf{Q}=\left(
\begin{array}{cccccc}
1 & 0 & \cdots & 0 & 0 & 0 \\
0 & 1 & \cdots & 0 & 0 & 0 \\
\vdots & \vdots & \ddots & \vdots & \vdots & \vdots\\
0 & 0 & \cdots & 1 & 0 & 0 \\
0 & 0 & \cdots & 0 & k & (1-k)
\end{array}
\right),
\end{equation}
with $0<k<1$. Then
\begin{equation}\label{SCEMM}
P = (P^1,...,P^m) \rightarrow \tilde{P} = (\tilde{P}^1,..., \tilde{P}^m, \tilde{P}^{m+1}) := (P^1,..., kP^m, (1-k)P^m).
\end{equation}
A vector in the tangent space of the simplex transforms in a similar way,
\begin{equation}
V = (V^1,...,V^m) \rightarrow \tilde{V} = (\tilde{V}^1,...,\tilde{V}^m, \tilde{V}^{m+1}) := (V^1,...,kV^m, (1-k)V^m).
\end{equation}

To prove uniqueness of the information metric, \v{C}encov \cite{C81} and Campbell \cite{C86} show that the only metric that preserves the inner product $<A,B>$ of two tangent vectors $A$, $B$ under a Markov mapping is precisely the information metric. It is straightforward to show that the metric has this property. To see this for the simple example discussed above, simply compute
\begin{eqnarray}
<\tilde{A},\tilde{B}> &=& \sum_{i=1}^{m+1} \left\{ \frac{\tilde{A}^i\,\tilde{B}^i}{\tilde{P}^i} \right\}
= \sum_{i=1}^{m-1} \left\{ \frac{A^i\,B^i}{P^i} \right\} + \frac{kA^m\,kB^m}{kP^m} + \frac{(1-k)A^m\,(k-1)B^m}{(1-k)P^m} \nonumber\\
&=& \sum_{i=1}^{m} \left\{ \frac{A^i\,B^i}{P^i} \right\} = <A,B>.
\end{eqnarray}
A proof of uniqueness is, as one would expect, not straightforward. The monograph of Caticha \cite{C12}, which includes a useful review of information geometry for both the discrete and continuous cases, provides a nice, accessible proof of uniqueness which follows the presentation of Campbell.

\section{Dynamics and symplectic geometry}

I now set the probabilities in motion. Assume that the time evolution of the $P^i$ is generated by an action principle and write the equations of motion using a Hamiltonian formalism. To do this, introduce additional coordinates $S^i$ which are canonically conjugate to the $P^i$ and a corresponding Poisson bracket for any two functions $F(P,S)$ and $G(P,S)$,
\begin{equation}\label{PoissonBrackets}
\left\{ F,G\right\} = \sum_i \left(
\frac{\partial F}{\partial P^i} \frac{\partial G}{\partial S^i}
 - \frac{\partial F}{\partial S^i}  \frac{\partial G}{\partial P^i} \right).
\end{equation}
As is well known, the Poisson bracket can be rewritten geometrically as
\begin{equation}
\left\{ F,G\right\} = \left( \partial F/\partial P\, , \; \partial F/\partial S\right) \, \Omega \, \left(
\begin{array}{c}
\partial G/\partial P \\
\partial G/\partial S
\end{array}
\right) ,
\end{equation}\nonumber
where $\Omega$ is the corresponding symplectic form, given in this case by
\begin{equation} \label{Omega}
\Omega=\left(
\begin{array}{cc}
0 & \textbf{1} \\
-\textbf{1} & 0
\end{array}
\right),
\end{equation}
where \textbf{1} is the $n \times n$ unit matrix. We thus have a symplectic structure and a corresponding {\it symplectic geometry\/}. The equations of motion for $P^i$ and $S^i$ are given by $\dot{P^i} = \left\{ P^i,H\right\}$, $\dot{S^i} = \left\{ S^i, H \right\}$ where ${H}$ is the Hamiltonian that generates time translations.

Notice that the 2$n$-dimensional phase space with coordinates $P^i$ and $S^i$ has a richer structure than the $n$-dimensional space of probabilities $P^i$; in particular, one may introduce the notion of {\it observables}, which are functions $O(P, S)$ of the coordinates, together with an algebra of observables defined in terms of the Poisson brackets of these functions \cite{H04}. However, not every function $O(P, S)$ qualifies as an observable because observables are also generators of infinitesimal transformations and these transformations must satisfy certain requirements. For example, the infinitesimal canonical transformation generated by any observable $O$ must preserve the normalization and positivity of $P$. This implies
\begin{equation}
O(P,S+\chi) = O(P,S),~~~~~\partial O / \partial S^i = 0 ~ \textrm{if} ~
P^i=0.
\end{equation}
Note that the first condition implies gauge invariance of the theory under $S^i \rightarrow S^i + \chi$, where $\chi$ is a constant \cite{H04}.

\section{K\"{a}hler geometry}

The 2$n$-dimensional phase space with coordinates $P^i$ and $S^i$ is an extension of the $n$-dimensional space of probabilities $P^i$. It is natural to ask the following question: Is it possible to extend the metric $G_{ij}$ in Eq.~(\ref{metricP}), which is only defined on the $n$-dimensional space of probabilities $P^i$, to the full $2n$-dimensional phase space of the $P^i$ and $S^i$? This can be done, but certain conditions which ensure the compatibility of the metric and symplectic structures have to be satisfied. These conditions are equivalent to requiring that the space have a K\"{a}hler structure (see the Appendix of Ref. \cite{RH12} for a proof).

A K\"{a}hler structure brings together metric, symplectic and complex structures in a harmonious way. To define such a space, introduce a complex structure $J_{\ b}^{a}$ and impose the following conditions \cite{G82},
\begin{eqnarray}
\Omega _{ab} &=& g_{ac}J_{\ b}^{c} \;,  \label{c1}\\
J_{\ c}^{a}g_{ab}J_{\ d}^{b} &=& g_{cd} \;,  \label{c2}\\
J_{\ b}^{a}J_{\ c}^{b} &=& -\delta _{\ c}^{a} \;.  \label{c3}
\end{eqnarray}
Eq. (\ref{c1}) is a compatibility equation between the symplectic structure $\Omega _{ab}$ and the metric $g_{ab}$, Eq. (\ref{c2}) is the condition that the metric should be Hermitian, and Eq. (\ref{c3}) is the condition that $J_{\ b}^{a}$ should be a complex structure.

The metric over the $n$-dimensional space of probabilities is the information metric, Eq. (\ref{metricP}). Then, metric over the full space will be of the form
\begin{equation}
g_{ab}=\left(
\begin{array}{cc}
\textbf{G} & \textbf{E} \\
\textbf{E}^{T} & \textbf{F}
\end{array}
\right),
\end{equation}
where $\textbf{G}=\texttt{diag}(\frac{\alpha}{2P^i})$, and \textbf{E} and \textbf{F} are $n \times n$ matrices that need to be determined.

A simple matrix calculation  using the K\"{a}hler conditions and the expression for $\Omega_{ab}$, Eq. (\ref{Omega}), leads to general forms for the metric $g_{ab}$ and the complex structure $J_{\ b}^{a}$,
\begin{equation}\label{gj_general}
g_{ab}=\left(
\begin{array}{cc}
\textbf{G} & \textbf{A}^{T} \\
\textbf{A} & (\textbf{1}+\textbf{A}^2)\textbf{G}^{-1}
\end{array}
\right),  ~~~~~
J_{\ b}^{a}=\left(
\begin{array}{cc}
\textbf{A} & (\textbf{1}+\textbf{A}^2)\textbf{G}^{-1} \\
-\textbf{G} & -\textbf{G}\textbf{A}\textbf{G}^{-1}
\end{array}
\right).
\end{equation}
where the $n \times n$ matrix \textbf{A} satisfies $\textbf{G}\textbf{A}\textbf{G}^{-1} =\textbf{A}^{T}$ but is otherwise arbitrary.

\section{On the geometry of the space of probabilities in motion}

At this point, it is useful to make some general remarks regarding the approach and the results obtained so far. The starting point is a system with a finite number of states (i.e., a discrete configuration space) and a probability $P=(P^1,...,P^n)$ where $n$ is the number of states. The first step of the procedure is to introduce the information metric, the natural metric on the space of probabilities. This leads to the most basic geometrical structure of the construction, \emph{information geometry}. The second step is to consider dynamics and to introduce an action principle to derive equations of motion. This is done using a Hamiltonian formalism: introduce coordinates $S^i$ canonically conjugate to the $P^i$, a Poisson bracket structure, and a Hamiltonian. This leads to additional geometrical structure, a \emph{symplectic structure}. The third step is to extend the metric structure of information geometry, to define a geometry over the full space of the $P^i$ and $S^i$. This can not be done in an arbitrary way. Consistency between the metric tensor and the symplectic form leads to a \emph{K\"{a}hler geometry}.

Notice that the construction is very general: It applies to {\it any} system with a finite number of states that is described probabilistically and which admits equations of motion that can be derived from an action principle. In particular, the construction does not require any assumptions regarding classical or quantum physics.

The few assumptions that enter into the analysis lead to the beautiful result that {\it the natural geometry of the space of probabilities in motion is a K\"{a}hler geometry\/}.

\section{Uniqueness of the K\"{a}hler metric via generalized Markov mappings}

The K\"{a}hler conditions impose strong restrictions on the form of the metric $g_{ab}$, Eq. (\ref{gj_general}), but that they do not determine the metric uniquely because it depends on a matrix $\textbf{A}$ which is to some extent arbitrary. Additional requirements are therefore needed to determine the form of $g_{ab}$.

As discussed before, Markov mappings play a crucial role in the proof of uniqueness of the information metric. In this section, it will be shown that the form of $\textbf{A}$ can be determined by requiring invariance of the metric $g_{ab}$ under a particular type of canonical transformation which extends the notion of a Markov mapping to the phase space with coordinates $P^i$ and $S^i$. The simplest way to introduce a generalization of a Markov mapping is to treat it as a point transformation (i.e., a transformation of the $P^i$) and to use the well known fact that a point transformation can always be extended to a canonical transformation.

Consider first the simple case of Markov mappings between spaces of the same dimension; i.e., where $n=m$. In this case, the Markov mappings are just permutations of the components $P^i$ and their extension to canonical transformations is trivial: carry out simultaneous permutations of the $P^i$ and $S^i$. These transformations have two important properties: they are linear and they do not mix the $P^i$ and the $S^i$.

The case that is non-trivial is the case where $n>m$. Here there are some subtle issues that need to be addressed, foremost that such canonical transformations will map spaces of different dimensions. This, however, is not a fundamental difficulty because there is a well developed theory of ``nonclassical canonical transformations'' which extends the concept of canonical transformations to allow for mappings of phase spaces of different dimensions \cite{S70, K77}. Although the canonical transformation that I derive in this section may be formulated within this formalism, a simpler approach is possible and therefore there will be no need to make use of the full theory of ``nonclassical canonical transformations.'' The simpler approach followed here consists of extending the dimensionality of the space in a trivial way and imposing constraints.

For the purpose of this paper, it will be sufficient to work out the generalization for the case of the simple example of a Markov mapping that I discussed before, where the dimensionality of the space of probabilities is increased by one; i.e., $n=m+1$. More generals cases can be derived by considering a series of successive transformations that are either permutations or which increase the dimensionality of the space by one at each step.

Consider then a system with states described by the coordinates $P^j,S^j$, $j=1,...m$, and a Hamiltonian $H(P^j,S^j)$ which describes the dynamics.

As a first step, increase the dimensionality of the space in a trivial way by adding coordinates $P^{m+1},S^{m+1}$ to the phase space. This increase in the dimensionality does not change the Hamiltonian. Therefore, the time evolution of the system remains the same and the additional coordinates $P^{m+1},S^{m+1}$ are {\it constants of the motion}. Consider now a point transformation relating old and new coordinates $P^k$, $\tilde{P}^k$ which satisfies the following relations,
\begin{eqnarray}\label{Ptransf}
P^i &=& \tilde{P}^i,\nonumber\\
P^m &=& \tilde{P}^m+\tilde{P}^{m+1},\nonumber\\
P^{m+1} &=& (1-k)\tilde{P}^m - k\tilde{P}^{m+1} \approx 0.
\end{eqnarray}
where $i=1,...,m-1$, and the symbol ``$\approx$'' is used to indicate a weak equality (i.e., a constraint in the sense of an equality of numerical values, not of functions of the phase space coordinates). Using the last two relations of Eqs. (\ref{Ptransf}), it is straightforward to show that $\tilde{P}^m$ and $\tilde{P}^{m+1}$ satisfy the constraints
\begin{eqnarray}
\tilde{P}^m &\approx& kP^m,\nonumber\\
\tilde{P}^{m+1} &\approx& (1-k)P^m,
\end{eqnarray}
which agree with Eq. (\ref{SCEMM}). Therefore, the relations defined in Eqs. (\ref{Ptransf}) are equivalent to a Markov mapping. Notice that these constraints are preserved because $P^{m+1}$ is a constant of the motion.

The second step is to extend this point transformation to a canonical transformation. In analogy to the case discussed above in which $n=m$, I will look for a linear canonical transformation which does not mix the $P^i$ and the $S^i$. Notice that these two conditions lead to a {\it unique} canonical transformation (up to additive constants which are unimportant). To define the canonical transformation, introduce the generating function
\begin{equation}
K = \sum_{i=1}^{m-1} \left\{ \tilde{P}^iS^i \right\} + (\tilde{P}^m + \tilde{P}^{m+1})\,S^m + [(1-k)\tilde{P}^m - k\tilde{P}^{m+1}]\,S^{m+1}.
\end{equation}
Derive the canonical transformation from the generating function in the standard way; i.e. $P^k = \partial K / \partial S^k$ and $\tilde{S}^k = \partial K / \partial \tilde{P}^k$. This leads to the following equations,
\begin{eqnarray}\label{CT}
P^i &=& \tilde{P}^i,~~~~~
P^m = \tilde{P}^m + \tilde{P}^{m+1},~~~~~~~~~~~~~~\;
P^{m+1} = (1-k) \tilde{P}^m - k \tilde{P}^{m+1},\nonumber\\
S^i &=& \tilde{S}^i,~~~~~
S^m = k\tilde{S}^m + (1-k)\tilde{S}^{m+1},~~~~~
S^{m+1} = \tilde{S}^m - \tilde{S}^{m+1},
\end{eqnarray}
and
\begin{eqnarray}\label{ICT}
\tilde{P}^i &=& P^i,~~~~~
\tilde{P}^m = kP^m + P^{m+1},~~~~~~~~~~~~~~
\tilde{P}^{m+1} = (1-k) P^m - P^{m+1},~~\nonumber\\
\tilde{S}^i &=& S^i,~~~~~
\tilde{S}^m = S^m + (1-k)S^{m+1},~~~~~~~~
\tilde{S}^{m+1} = S^m - k S^{m+1}.
\end{eqnarray}
where $i=1,...,m-1$.

I have increased the dimensionality of the phase space by two dimensions and I have two constants of the motion, $P^{m+1}$ and $S^{m+1}$. As shown in \ref{AppCT}, consistency requires both $P^{m+1} \approx 0$ and $S^{m+1} \approx 0$, which leads to
\begin{eqnarray}\label{ConstPrimed}
\tilde{P}^m &\approx& kP^m,~~~~~\tilde{P}^{m+1} \approx (1-k)P^m,\nonumber\\
\tilde{S}^m &\approx& S^m,~~~~~~~\tilde{S}^{m+1} \approx S^m.
\end{eqnarray}
It is also shown in \ref{AppCT} that the time evolution is preserved; i.e., the dynamics in the $2(n+1)$-dimensional phase space (with coordinates with tildes) reproduces precisely the dynamics in the $2n$-dimensional phase space (with coordinates without tildes).

Given the canonical transformation, Eq. (\ref{CT}), and its inverse, Eq. (\ref{ICT}), one can examine the restrictions imposed on the matrix $\textbf{A}$ by the requirement of invariance of the metric $g_{ab}$ under this generalization of a Markov mapping. The calculation is summarized in \ref{AppGCEMM}. The result is that the matrix $\textbf{A}$ must be proportional to the $n \times n$ unit matrix, $\textbf{A}=A\textbf{1}$, where $A$ is a constant. The line element depends on two parameteres only, $\alpha$ and $A$, and it takes the remarkably simple form
\begin{equation}\label{FlatKaehlerLineElementPS}
dl^2 = \sum_k \left[ \frac{\alpha}{2P^k}(dP^k)^2 + 2 A dP^k dS^k + \frac{2P^k}{\alpha}(1+A^2)(dS^k)^2 \right].
\end{equation}

\section{Complex coordinates and wave functions}

Up to now, I have made use of real coordinates ${P^i}$, ${S^i}$. K\"{a}hler geometry, however, is best expressed in terms of complex coordinates. I carry out a complex transformation that shows that the metric of Eq. (\ref{FlatKaehlerLineElementPS}) describes in fact a flat K\"{a}hler space.

Set $\textbf{A}=A\textbf{1}$ in Eqs. (\ref{gj_general}) and consider first the particular case $A=0$. The tensors that define the K\"{a}hler structure take the form
\begin{equation}
\Omega _{ab}=\left(
\begin{array}{cc}
0 & 1 \\
-1 & 0
\end{array}
\right),  ~
g_{ab}=\left(
\begin{array}{cc}
\textbf{G} & 0\\
0 & \textbf{G}^{-1}
\end{array}
\right),  ~
J_{\ b}^{a}=\left(
\begin{array}{cc}
0 & \textbf{G}^{-1} \\
-\textbf{G} & 0
\end{array}
\right).
\end{equation}
Define now the {\it Madelung transformation},
\begin{equation}\label{MT}
\psi^k =\sqrt{P^k}\exp (iS^k/\alpha),~~~~~~~~~~\bar{\psi}^k=\sqrt{P^k}\exp (-iS^k/\alpha).
\end{equation}
A simple calculation shows that the tensors that define the K\"{a}hler geometry, expressed in terms of $\psi^k$, $\bar{\psi}^k$, take the standard form which is characteristic of flat-space \cite{G82},
\begin{equation}\label{OgJpsipsistar}
\Omega _{ab}=\left(
\begin{array}{cc}
0 & i\alpha \textbf{1} \\
-i\alpha \textbf{1} & 0
\end{array}
\right),~
g_{ab}=\left(
\begin{array}{cc}
0 &  \alpha \textbf{1} \\
\alpha \textbf{1} & 0
\end{array}
\right),~
J_{\ b}^{a}=\left(
\begin{array}{cc}
-i \textbf{1} & 0 \\
0 & i \textbf{1}
\end{array}
\right).
\end{equation}
One may conclude that in this case ($A=0$) there is a natural set of fundamental variables given by $\psi^i$ and $\bar{\psi}^i$. In terms of these variables, the tensors that define the K\"{a}hler geometry take their simplest form. If the constant $\alpha$ is set equal to $\hbar$, these fundamental variables are precisely the {\it wave functions\/} of quantum mechanics. This is a remarkable result because it is based on geometrical arguments only. The derivation does not use any assumptions from quantum theory.

Consider now the more general case $A \neq 0$. The tensors that define the K\"{a}hler structure take the form
\begin{equation}
\Omega _{ab}=\left(
\begin{array}{cc}
0 & 1 \\
-1 & 0
\end{array}
\right),  ~
g_{ab}=\left(
\begin{array}{cc}
\textbf{G} & A \, \textbf{1} \\
A \, \textbf{1} & (1+A^2)\textbf{G}^{-1}
\end{array}
\right),  ~
J_{\ b}^{a}=\left(
\begin{array}{cc}
A \, \textbf{1} & (1+A^2)\textbf{G}^{-1} \\
-\textbf{G} & -A \, \textbf{1}
\end{array}
\right).
\end{equation}
In this case, define the {\it modified Madelung transformation}
\begin{equation}\label{MMT}
\phi^k = \sqrt{P^k}\exp\left[i \left(\Lambda \, S^k / \alpha  - \gamma \ln \sqrt{P^k} \right) \right],~~~~~\bar{\phi}^k = \sqrt{P^k}\exp\left[-i \left(\Lambda \, S^k / \alpha  - \gamma \ln \sqrt{P^k} \right) \right],
\end{equation}
where $\Lambda = 1/(1+A^2)$ and $\gamma =  -A/(1+A^2)$. Once more, the tensors that define the K\"{a}hler geometry, expressed now in terms of $\phi^k$, $\bar{\phi}^k$, take the standard form which is characteristic of flat-space,
\begin{equation}
\Omega _{ab}=\left(
\begin{array}{cc}
0 & i\alpha \Lambda^{-1} \textbf{1} \\
-i\alpha \Lambda^{-1} \textbf{1} & 0
\end{array}
\right),~
g_{ab}=\left(
\begin{array}{cc}
0 &  \alpha \Lambda^{-1} \textbf{1} \\
\alpha \Lambda^{-1} \textbf{1} & 0
\end{array}
\right),~\nonumber\\
J_{\ b}^{a}=\left(
\begin{array}{cc}
-i \textbf{1} & 0 \\
0 & i \textbf{1}
\end{array}
\right).
\end{equation}
This shows that the geometry of the K\"{a}hler space is the same whether $A=0$ or $A\neq0$. In fact, it is possible to map one case to the other using an $A$-dependent canonical transformation. It is clear then that both cases lead to the same theory (provided one sets $\alpha=\hbar$ when $A=0$ or $\alpha \Lambda^{-1}=\hbar$ when $A\neq0$), and in the following sections I will set $A=0$ and use the complex coordinates (wave functions) $\psi^i$ and $\bar{\psi}^i$.

The transformation that takes you from the coordinates of Eq. (\ref{MMT}) to the coordinates of Eq. (\ref{MMT}) is a particular case of a family of nonlinear gauge transformations introduced by Doebner and Goldin \cite{DG96} (compare to their Eq. (2.2)). As pointed out by Doebner and Goldin, the theory that results from this particular family of nonlinear gauge transformations is physically equivalent to standard quantum mechanics. Here we arrive at the same conclusion, but now on the basis of the equivalence of the two cases $A=0$ and $A\neq0$ via a canonical transformation. One may therefore view the present derivation of the geometric formulation of quantum mechanics as providing a new route to this family of Doebner-Goldin nonlinear gauge transformations.

\section{Group of unitary transformations and Hilbert space formulation}

I now show that the group of transformations of the theory is the unitary group and that one may introduce a Hilbert space formulation. Both of these results are needed to establish the equivalence of the geometric formulation derived here to standard quantum mechanics.

Since the K\"{a}hler structure includes a symplectic structure, the group of symplectic transformations, Sp($2n$,$R$), will play an important role in the theory. But the group of transformations of the theory can not be the symplectic group because the transformations have to satisfy certain requirements. The first requirement is that they preserve the normalization of the probability, $\sum_{i}P^i = \sum_{i}\psi^i \bar{\psi}^i = 1$. The second requirement is that the metric be form invariant under the transformations; i.e., that the line element $dl^2 = 2 \alpha \sum_j d\bar{\psi}^j d\psi^j$ of the K\"{a}hler space be preserved by the transformations. Requiring normalization of the probability and metric invariance leads to the group of rotations on the $2n$-dimensional sphere, O($2n,R$).

Unitary transformations are the only symplectic transformations which are also rotations; i.e., Sp($2n$,$R$) $\cap$ O($2n,R$)$=U(n)$ \cite{A78}. Therefore, the group of transformations of the theory is precisely the group of unitary transformations.

One can now introduce a Hilbert space formulation. There is a standard construction that associates a complex Hilbert space with any K\"{a}hler space. Given two complex vectors $\psi^i$ and $\varphi^i$, define the Dirac product by \cite{K79}
\begin{eqnarray}
\langle \psi |\varphi \rangle &=&\frac{1}{2 \alpha}\sum_i \left\{ \left(
\psi^i,\bar{\psi}^i \right) \cdot \left[
g+i\Omega \right] \cdot \left(
\begin{array}{c}
\varphi^i \nonumber\\
\bar{\varphi}^i
\end{array}
\right) \right\}  \\
&=&\frac{1}{2 }\sum_i \left\{ \left( \psi^i,\bar{\psi}^i \right) \left[ \left(
\begin{array}{cc}
0 & \textbf{1} \\
\textbf{1} & 0
\end{array}
\right) +i\left(
\begin{array}{cc}
0 & i \textbf{1} \\
-i \textbf{1} & 0
\end{array}
\right) \right] \left(
\begin{array}{c}
\varphi^i \nonumber\\
\bar{\varphi}^i
\end{array}
\right) \right\}  \nonumber\\
&=& \sum_i \bar{\psi}^i \varphi^i
\end{eqnarray}
In this way one arrives at the Hilbert space formulation of quantum mechanics.

This suggests that the Hilbert space structure of quantum mechanics is perhaps not as fundamental as its geometrical structure.

\section{Concluding remarks}

The geometry of quantum theory can be derived from information geometry, the natural geometry on the space of probabilities, using only a few assumptions. The derivation has a number of interesting features:
\begin{itemize}
 \item {Doubling of the dimensionality of the space} (i.e., $\{P^i\} \rightarrow \{P^i,S^i\}$) from dynamical considerations,
 \item {Complex structure} from consistency between metric and symplectic structures,
 \item {Wave functions} as the natural complex coordinates of the K\"{a}hler space,
 \item Representation in terms of canonical transformations of a particular case of a family of Doebner-Goldin nonlinear gauge transformations,
 \item {Unitary transformations} as the group of transformations allowed by the theory,
 \item {Hilbert space formulation} expressed in terms of geometrical quantites associated with the K\"{a}hler space.
\end{itemize}

The derivation presented here relies heavily on, and extends, a geometrical reconstruction of quantum theory by Reginatto and Hall which takes information geometry as its starting point \cite{RH12,RH13}. Mehrafarin \cite{M05} and Goyal \cite{G08, G10} have also developed reconstructions of quantum theory using information-geometrical approaches. A detailed comparison to their approaches has not been carried out yet; however, one of the main differences is in the handling of {\it dynamics}, which plays a crucial role here. In particular, the use of an action principle to describe the dynamics of probabilities leads in a natural way to geometrical structure that goes beyond information geometry. A generalization of Markov mappings has proven to be very useful for deriving a unique K\"{a}hler geometry; these types of transformations may be of interest for other classes of problems where probabilities play an important role.

\section{Acknowledgments}

I am very grateful to the organizers of the Symmetries in Science XVI symposium for the invitation to attend a wonderful and stimulating meeting. I also want to thank Michael J. W. Hall, Ariel Caticha, Gerald A. Goldin and Heinz-Dietrich Doebner for their comments and suggestions regarding the work presented in this paper.

\appendix

\section{Generalized Markov mappings: constants of the motion and dynamics}\label{AppCT}

To derive the canonical transformation that generalizes the Markov mapping of Eq. (\ref{SCEMM}), the dimensionality of the original phase space was increased by two in a trivial way. This led to two constants of the motion, $P^{m+1}$ and $S^{m+1}$. $P^{m+1}$ was set to $P^{m+1} \approx 0$, with corresponding constraints for the $\tilde{P}^k$ of the form
\begin{eqnarray}
\tilde{P}^m &\approx& kP^m,\nonumber\\
\tilde{P}^{m+1} &\approx& (1-k)P^m.
\end{eqnarray}

These are precisely the conditions that are needed to get a generalization of the Markov mapping of Eq. (\ref{SCEMM}). When $k=1/2$, $\tilde{P}^m \approx \tilde{P}^{m+1}$, which is expected because in this case there should be invariance under the re-labeling $m \leftrightarrow m+1$. To fix the value of $S^{m+1}$, notice that $S^{m+1} \approx c$ leads to constraints for the $\tilde{S}^k$ of the form
\begin{eqnarray}
\tilde{S}^m &\approx& S^m+(1-k) c,\nonumber\\
\tilde{S}^{m+1} &\approx& S^m-k c.
\end{eqnarray}
Argue once more that there should be invariance under the re-labeling $m \leftrightarrow m+1$ in the case when $k=1/2$. But this can only be satisfied if $c=0$. On can conclude therefore that the constants of the motion must satisfy
\begin{eqnarray}
P^{m+1} &=& (1-k) \tilde{P}^m - k \tilde{P}^{m+1} \approx 0,\nonumber\\
S^{m+1} &=& \tilde{S}^m - \tilde{S}^{m+1} \approx 0.
\end{eqnarray}
The corresponding constraints for the unprimed coordinates are of the form
\begin{eqnarray}
\tilde{P}^m &\approx& kP^m,~~~~~\tilde{P}^{m+1} \approx (1-k)P^m,\nonumber\\
\tilde{S}^m &\approx& S^m,~~~~~~~\tilde{S}^{m+1} \approx S^m.
\end{eqnarray}

I check now that the dynamics in the $2(n+1)$-dimensional phase space (with coordinates with tildes) reproduces precisely the dynamics in the original $2n$-dimensional phase space (with coordinates without tildes). Using Eqs. (\ref{CT}-\ref{ICT}), one can show that
\begin{eqnarray}
\dot{\tilde{P}}^i &=& \frac{\partial H}{\partial S^i},~~~~~~~~
\dot{\tilde{P}}^m = \frac{\partial H}{\partial S^m}\,k,~~~~~
\dot{\tilde{P}}^{m+1} = \frac{\partial H}{\partial S^m}\,(1-k),\nonumber\\
\dot{\tilde{S}}^i &=& -\frac{\partial H}{\partial P^i},~~~~~
\dot{\tilde{S}}^m = -\frac{\partial H}{\partial P^m},~~~~~
\dot{\tilde{S}}^{m+1} = - \frac{\partial H}{\partial P^m}.
\end{eqnarray}
These equations lead to
\begin{eqnarray}
\dot{P}^i &=& \dot{\tilde{P}}^i = \frac{\partial H}{\partial S^i},~~~~~~~~
\dot{P}^m = \dot{\tilde{P}}^m + \dot{\tilde{P}}^{m+1} = \frac{\partial H}{\partial S^m}\nonumber\\
\dot{S}^i &=& \dot{\tilde{S}}^i = -\frac{\partial H}{\partial P^i},~~~~~
\dot{S}^m = k\,\dot{\tilde{S}}^m + (1-k)\,\dot{\tilde{S}}^{m+1} = -\frac{\partial H}{\partial P^m},
\end{eqnarray}
which are the correct equations of motion for the original space.

\section{Invariance of the K\"{a}hler metric under generalized Markov mappings}\label{AppGCEMM}

The metric of the K\"{a}hler space is given by
\begin{equation}
g_{ab}=\left(
\begin{array}{cc}
\textbf{G} & \textbf{A}^{T} \\
\textbf{A} & (\textbf{1}+\textbf{A}^2)\textbf{G}^{-1}
\end{array}
\right),
\end{equation}
where $\textbf{G}=\texttt{diag}(\frac{\alpha}{2P^i})$ and the $n \times n$ matrix \textbf{A} satisfies $\textbf{G}\textbf{A}\textbf{G}^{-1} =\textbf{A}^{T}$. For the calculations in this Appendix it is convenient to introduce the matrix $\textbf{B}$ with matrix elements given by
\begin{equation}
B_{jk}=\sqrt{{P_j}/{P_k}}\;A_{jk}.
\end{equation}
It is straightforward to show that $\textbf{B}$ is a symmetric matrix, $B_{jk}=B_{kj}$.

To restrict the form of $\textbf{B}$, it will be sufficient to consider the invariance of the metric under the particular generalized Markov mapping which corresponds to the inverse canonical transformation of Eq. (\ref{ICT}). After taking into consideration the constraints, Eq. (\ref{ConstPrimed}), the generalized Markov mapping can be written in the form
\begin{eqnarray}\label{B3}
P &=& (P^i,P^m) \rightarrow~~ \tilde{P} = (\tilde{P}^i, \tilde{P}^m, \tilde{P}^{m+1}) := (P^i, kP^m, (1-k)P^m),\nonumber\\
S &=& (S^i,S^m)\, \rightarrow~~ \tilde{S} = (\tilde{S}^i, \tilde{S}^m, \tilde{S}^{m+1}) ~\,:= (S^i, S^m, S^m),
\end{eqnarray}
where $i=1,...,m-1$.

As a first step, look at the contribution to the line element $dl^2$ from the mixed terms $dP^k dS^k$. In terms of the coordinates without tildes,
\begin{equation}\label{B4}
dl^2 = \sum_{i=1}^{m-1}\left\{ B_{ii}dP^i dS^i + \sqrt{\frac{P^i}{P^m}}B_{im}dP^i dS^m + \sqrt{\frac{P^m}{P^i}}B_{mi}dP^m dS^i \right\} + B_{mm}dP^m dS^m.
\end{equation}
There is a corresponding expression for the coordinates with tildes, and with the help of Eq. (\ref{B3}) it can be rewritten in terms of coordinates without tildes. This leads to
\begin{eqnarray}\label{B5}
dl^2 &=& \sum_{i=1}^{m-1}\left\{ \tilde{B}_{ii}dP^i dS^i + \left[\sqrt{\frac{P^i}{kP^m}}\tilde{B}_{im} + \sqrt{\frac{P^i}{(1-k)P^m}}\tilde{B}_{i(m+1)}\right]dP^i dS^m \right\} \nonumber\\
&~&+ \sum_{i=1}^{m-1}\left\{ \left[\sqrt{\frac{k^3P^m}{P^i}}\tilde{B}_{mi} + \sqrt{\frac{(1-k)^3P^m}{P^i}}\tilde{B}_{(m+1)i}\right]dP^m dS^i \right\} \nonumber\\
&~&+ \left[k\tilde{B}_{mm} + \sqrt{\frac{k^3}{1-k}}\tilde{B}_{m(m+1)}  + \sqrt{\frac{(1-k)^3}{k}}\tilde{B}_{(m+1)m} + (1-k)\tilde{B}_{(m+1)(m+1)}\right]dP^m dS^m.\nonumber\\
\end{eqnarray}
Equate terms in Eqs. (\ref{B4}) and (\ref{B5}) proportional to the same $dP^a dS^b$, where $a,b=1,...,m$. This leads to the four relations
\begin{eqnarray}
B_{ii} &=& \tilde{B}_{ii}, \nonumber\\
B_{im} &=& \sqrt{\frac{1}{k}}\tilde{B}_{im} + \sqrt{\frac{1}{(1-k)}}\tilde{B}_{i(m+1)}, \nonumber\\
B_{mi} &=& \sqrt{k^3}\tilde{B}_{mi} + \sqrt{(1-k)^3}\tilde{B}_{(m+1)i}, \nonumber\\
B_{mm} &=& k\tilde{B}_{mm} + \sqrt{\frac{k^3}{1-k}}\tilde{B}_{m(m+1)}  + \sqrt{\frac{(1-k)^3}{k}}\tilde{B}_{(m+1)m} + (1-k)\tilde{B}_{(m+1)(m+1)}.
\end{eqnarray}
Since the matrix $\textbf{B}$ is symmetric, $B_{im}=B_{mi}$, which leads to
\begin{equation}
\sqrt{\frac{1}{k}}\tilde{B}_{im} + \sqrt{\frac{1}{(1-k)}}\tilde{B}_{i(m+1)} = \sqrt{k^3}\tilde{B}_{mi} + \sqrt{(1-k)^3}\tilde{B}_{(m+1)i}.
\end{equation}
By symmetry, $\tilde{B}_{im}=\tilde{B}_{i(m+1)}$ at $k=1/2$, but this relation can only be satisfied if $\tilde{B}_{im}=\tilde{B}_{i(m+1)}=0$, which in turn implies $B_{im}=B_{mi}=0$. Since $B_{im}$ and $B_{mi}$ are independent of $k$, it follows that they must always be zero. This shows that the off-diagonal elements of the matrix \textbf{B} are zero.

Now look at the contribution to the line element $dl^2$ from terms proportional to $dP^a dP^a$ and $dS^a dS^a$. The terms proportional to $dP^a dP^a$ give the two relations
\begin{eqnarray}\label{B8}
B_{ii} &=& \tilde{B}_{ii}, \nonumber\\
B_{mm} &=& k\tilde{B}_{mm} + (1-k)\tilde{B}_{(m+1)(m+1)},
\end{eqnarray}
while the terms proportional to $dS^a dS^a$ give the two relations
\begin{eqnarray}\label{B9}
1+B_{ii}^2 &=& 1+\tilde{B}_{ii}^{~2}, \nonumber\\
1+B_{mm}^2 &=& k(1+\tilde{B}_{mm}^{~2})+(1-k)\left(1+\tilde{B}_{(m+1)(m+1)}^{~2}\right).
\end{eqnarray}
Combining Eqs. (\ref{B8}) and (\ref{B9}) leads to
\begin{eqnarray}
B_{ii} &=& \tilde{B}_{ii}, \nonumber\\
B_{mm} &=& \tilde{B}_{mm}= \tilde{B}_{(m+1)(m+1)}.
\end{eqnarray}
Notice that this result is valid for arbitrary values of $k$. Since there is nothing special about the particular labels $m$ and $(m+1)$, {\it all} the diagonal elements of the matrices $\textbf{B}$ and $\tilde{\textbf{B}}$ must be equal. Then
\begin{eqnarray}
\textbf{B} &=& B \textbf{1}_{m \times m}, \nonumber\\
\tilde{\textbf{B}} &=& B \textbf{1}_{(m+1) \times (m+1)},
\end{eqnarray}
where $\textbf{1}_{n \times n}$ is the $n \times n$ unit matrix and $B$ still has to be determined.

To carry out this last step, use the relations
\begin{eqnarray}
B(P,S) &=& B(P^i,P^m,S^i,S^m), \nonumber\\
\tilde{B}(\tilde{P},\tilde{S}) &=& B(P(\tilde{P}),S(\tilde{S})) = B(\tilde{P}^i,\tilde{P}^m+\tilde{P}^{m+1},\tilde{S}^i,k\tilde{S}^m+(1-k)\tilde{S}^{m+1}).
\end{eqnarray}
The functional form of $B(P,S)$ must be the same as the functional form of $\tilde{B}(\tilde{P},\tilde{S})$, and these expressions must be both invariant under permutations and independent of $k$. The only functional form that seems to satisfy all these conditions appears to be $B(P,S)=B(\sum_i P^i)$. But $\sum_i P^i=1$, therefore one can conclude that $B$ is a {\it constant} and the matrix
\begin{equation}
\textbf{B}=B \textbf{1}
\end{equation}
is a constant matrix proportional to the unit matrix. This in turn implies that
\begin{equation}
\textbf{A}=A \textbf{1}
\end{equation}
where $A$ is a constant.


\section*{References}
\begin{thebibliography}{9}

\bibitem{K79} Kibble T W B 1979 {\it Commun. math. Phys.} \textbf{65} 189

\bibitem{RH12} Reginatto M and Hall M J W 2012 {\it AIP Conf. Proc.} \textbf{1443} 96

\bibitem{RH13} Reginatto M and Hall M J W 2013 {\it AIP Conf. Proc.} \textbf{1553} 246

\bibitem{B43} Bhattacharyya A 1943 {\it Bull. Calcutta Math. Soc.} \textbf{35} 99

\bibitem{B46} Bhattacharyya A 1946 {\it Sanhky\={a}} \textbf{7} 401

\bibitem{G90} Good I J 1990 {\it J. Stat. Comput. Simulation} \textbf{36} 179

\bibitem{W81} Wootters W K 1981 {\it Phys. Rev. D.} \textbf{23} 357

\bibitem{C81} \v{C}encov N N 1981 {\it Statistical decision rules and optimal inference} (Transl. Math. Monographs \textbf{53}) (Providence, R. I.: Amer. Math. Soc.)

\bibitem{C86} Campbell L L 1986 {\it Proc. Amer. Math. Soc.} \textbf{98} 135

\bibitem{C12} Caticha A 2012 {\it Entropic Inference and the Foundations of Physics} (Sao Paulo, Brazil: Brazilian Chapter of the International Society for Bayesian Analysis -- ISBrA)

\bibitem{H04} Hall M J W 2004 {\it J. Phys. A} \textbf{37} 7799

\bibitem{G82} Goldberg S I 1982 {\it Curvature and Homology} (New York: Dover Publications)

\bibitem{S70} Scheifele G 1970 {\it Celest. Mech.} \textbf{2} 296

\bibitem{K77} Kurcheeva I V 1977 {\it Celest. Mech.} \textbf{15} 353

\bibitem{DG96} Doebner H-D and Goldin G A 1996 {\it Phys. Rev. A} \textbf{54}, 3764

\bibitem{A78} Arnold V I 1978 {\it Mathematical methods of classical mechanics} (Berlin: Springer)

\bibitem{M05} Mehrafarin M 2005 {\it	Int. J. Theor. Phys.} \textbf{44} 429

\bibitem{G08} Goyal P 2008 {\it Phys. Rev. A} \textbf{78} 052120

\bibitem{G10} Goyal P 2010 {\it New Journal of Physics} \textbf{12} 023012

\end{thebibliography}
\end{document}